\documentclass{article}
\usepackage[labelfont=bf]{caption} %
\usepackage[font={small}]{caption}   %

\usepackage{graphicx}%
\usepackage{dcolumn}%
\usepackage{bm}%
\usepackage[pdftex,colorlinks=true,linkcolor=blue,citecolor=blue]{hyperref}
\usepackage[utf8]{inputenc}
\usepackage{lipsum}
\usepackage{newunicodechar}
\newunicodechar{−}{\ensuremath{-}}
\usepackage{amsmath}
\usepackage{authblk}
\usepackage{amsfonts}
\usepackage{placeins}
\usepackage[numbers, sort&compress]{natbib}  %
\usepackage{bibunits}
\usepackage{amssymb}
\usepackage{pdflscape} %
\usepackage{multirow}
\usepackage{xcolor}

\usepackage[table, dvipsnames]{xcolor}
\usepackage{makecell}
\usepackage{boldline}
\setcellgapes{3pt}
\usepackage{amsmath}
\usepackage{amssymb}

\usepackage{textgreek}      %

\usepackage[strict]{changepage} %
\usepackage{nicefrac}

\defaultbibliographystyle{naturemag}
\defaultbibliography{references}

    \usepackage{soul}   %
    \setstcolor{red} %
    \setul{}{1.5pt}  %

\usepackage{lineno}

\usepackage{paralist}

\usepackage{xspace}

\title{An integrated multi-THz tunable linear isolator based on electro-optic non-reciprocal strong coupling}

\author{
    Gwan In Kim$^1$, Violet Workman$^2$, Oğulcan E. Örsel$^1$, Jieun Yim$^3$, and Gaurav Bahl$^{3*}$ \\
    $^1$ Department of Electrical $\&$ Computer Engineering, \\
    $^2$ Department of Physics, \\
    $^3$ Department of Mechanical Science $\&$ Engineering, \\
    University of Illinois at Urbana–Champaign, Urbana, IL 61801 USA \\
}

\date{}

\begin{document}
\begin{bibunit}

\maketitle


\begin{abstract}
Optical isolators are essential for laser protection and robust signal routing, but the incorporation of the necessary magneto-optic (MO) materials in foundries has remained a challenge. As an alternative, several integrated non-magnetic isolators based on acousto-optic (AO) and electro-optic (EO) spatio-temporal modulation have been proposed. Unlike MO isolators, these solutions are wavelength agnostic, though few published demonstrations reach performance that is comparable to MO devices.
The most significant remaining concerns are on mitigating undesirable sidebands, 
achieving wide bandwidth or wide tunability, and having a design that is practical to deploy.
Most of these challenges can be addressed through \mbox{\it non-reciprocal} strong coupling between waveguide or resonator modes, with the intent to produce extremely asymmetric optical dispersion, but this has never been accomplished with electro-optics.
Here we demonstrate a compact EO optical isolator, using thin film lithium niobate, that is the first EO device to reach the non-reciprocal strong coupling regime. In this new regime, the isolator produces a very high isolation figure of merit ($>32$ dB contrast per dB of insertion loss, 47.7 dB isolation contrast with 1.45 dB insertion loss)
and, due to its architecture, achieves linear operation with negligible sideband generation. We additionally demonstrate THz-scale (8 nm) tunability of the isolation band that is not fundamentally limited, and can be extended to multi-THz operation. 

\end{abstract}

\vspace{12pt}
Magneto-optic isolators have persisted as the dominant non-reciprocal technology. However, there remain unsolved challenges with bringing the necessary magneto-optic materials onto integrated photonics platforms~\cite{bi_-chip_2011,ghosh_ceyigsilicon--insulator_2012,huang_dynamically_2017,zhang_monolithic_2019,du_monolithic_2018, yan_waveguide-integrated_2020, yan_ultra-broadband_2024}, primarily due to foundry constraints, but also due to the requirement of magnetic biasing, and the large absorption and strong chromatic dependence of the magneto-optic materials.
In recent years, there have emerged alternative electrically-driven non-reciprocal systems that use either acousto-optic (AO)~\cite{hwang_all-fiber-optic_1997,sohn_time-reversal_2018,kittlaus_electrically_2020,sohn_electrically_2021,tian_magnetic-free_2021,cheng_terahertz-bandwidth_2025} or electro-optic (EO)~\cite{bhandare_novel_2005, doerr_optical_2011,lira_electrically_2012, tzuang_non-reciprocal_2014, doerr_silicon_2014, dostart_optical_2021, yu_integrated_2023, orsel_electro-optic_2023, gao_thin-film_2024, orsel_giant_2025} symmetry breaking to produce a direction sensitive optical response. The general principle behind these devices is to generate a traveling wave index modulation, or spatio-temporal modulation, from which light experiences some combination of non-reciprocal operations.
Non-magnetic isolators that use these approaches (Supplementary Data Table~\ref{tab:FoM_table}) have been able to exhibit wide bandwidth non-reciprocal effects~\cite{doerr_silicon_2014, yu_integrated_2023, gao_thin-film_2024,cheng_terahertz-bandwidth_2025}, very low insertion loss~\cite{sohn_electrically_2021, tian_magnetic-free_2021, yu_integrated_2023}, ultra-high contrast~\cite{bhandare_novel_2005, tian_magnetic-free_2021, sohn_electrically_2021, yu_integrated_2023, gao_thin-film_2024}, and linear response~\cite{sohn_electrically_2021, tian_magnetic-free_2021,cheng_terahertz-bandwidth_2025}, though currently there is no single solution that delivers on all of these metrics simultaneously.

A broad category of non-magnetic isolators leverage non-reciprocal interference induced by spatio-temporal modulation \cite{bhandare_novel_2005, doerr_optical_2011, tzuang_non-reciprocal_2014, dostart_optical_2021, yu_integrated_2023, gao_thin-film_2024, orsel_giant_2025}. These devices are relatively easy to produce, but they require precise tuning of the amplitude and phase of the electrical stimulus for best performance, and also generate a very large number of spurious sidebands.
The second broad category of non-magnetic isolators leverage non-reciprocal inter-modal conversion~\cite{hwang_all-fiber-optic_1997,lira_electrically_2012, sohn_time-reversal_2018, sohn_electrically_2021, tian_magnetic-free_2021, kittlaus_electrically_2020,cheng_terahertz-bandwidth_2025}, also driven by spatio-temporal modulation. These devices do not require precise control of the electrical stimuli, as the non-reciprocity is determined by the availability of specific optical modes and not through interference effects. Additionally, all undesirable sidebands are suppressed due to the absence of suitable optical states. However, these devices require stringent phase matching in both frequency and momentum space which increases the design complexity.

A particularly effective version of the inter-modal approach is the chiral shunt design demonstrated by Sohn et al.~{\cite{sohn_electrically_2021}} and the similar design by Tian et al.~{\cite{tian_magnetic-free_2021}}, which can simultaneously achieve ultra-low insertion loss and large directional contrast, with linearity (minimal frequency shifted sidebands) and linear operation (input-output relation). However, since this design relies on AO inter-modal coupling, the phase-matching is very strict and the isolation is both narrowband and difficult to spectrally tune. This is because the acoustic dispersion locks the frequency and momentum of the acoustic stimulus, and because acoustic transducers provide only a limited driving bandwidth. These AO designs also require power to be converted from an input RF electrical stimulus into acoustic domain, and the acoustic waves subsequently propagate away from the device or are dissipated.

In this work we experimentally demonstrate a compact EO isolator that brings transformative improvements to the chiral shunt design -- significantly relaxing the phase matching, achieving multi-THz tunability, and enabling protective oxide cladding. 
We demonstrate tuning of this isolator over a wide \mbox{8 nm} span, that is only limited by our experimental capability, and we confirm that it can consistently maintain very large isolation contrast (IC $>41$ dB) with very low insertion loss (IL $<2$ dB) over this tuning range. Our best result shows IC $=47.7$ dB simultaneously with IL $=1.45$ dB.
In doing so, we also achieve the first EO device to operate in the non-reciprocal strong coupling regime.
There are three major design innovations that enable this result. First, RF excitations have high phase velocity and do not naturally carry sufficient momentum to allow for the required phase-matching, which we resolve here with a synthetic momentum approach. Second, we simplify the optical mode placement in frequency-momentum space by means of a double racetrack resonator that allows for flexible momentum spacing by design. Finally, we implement an EO electrode set that can successfully break the orthogonality between these optical modes while achieving a large coupling rate.

\begin{figure}[b!]
    \begin{adjustwidth*}{-1in}{-1in}
    \hsize=\linewidth
    \includegraphics[width=1.2\columnwidth]{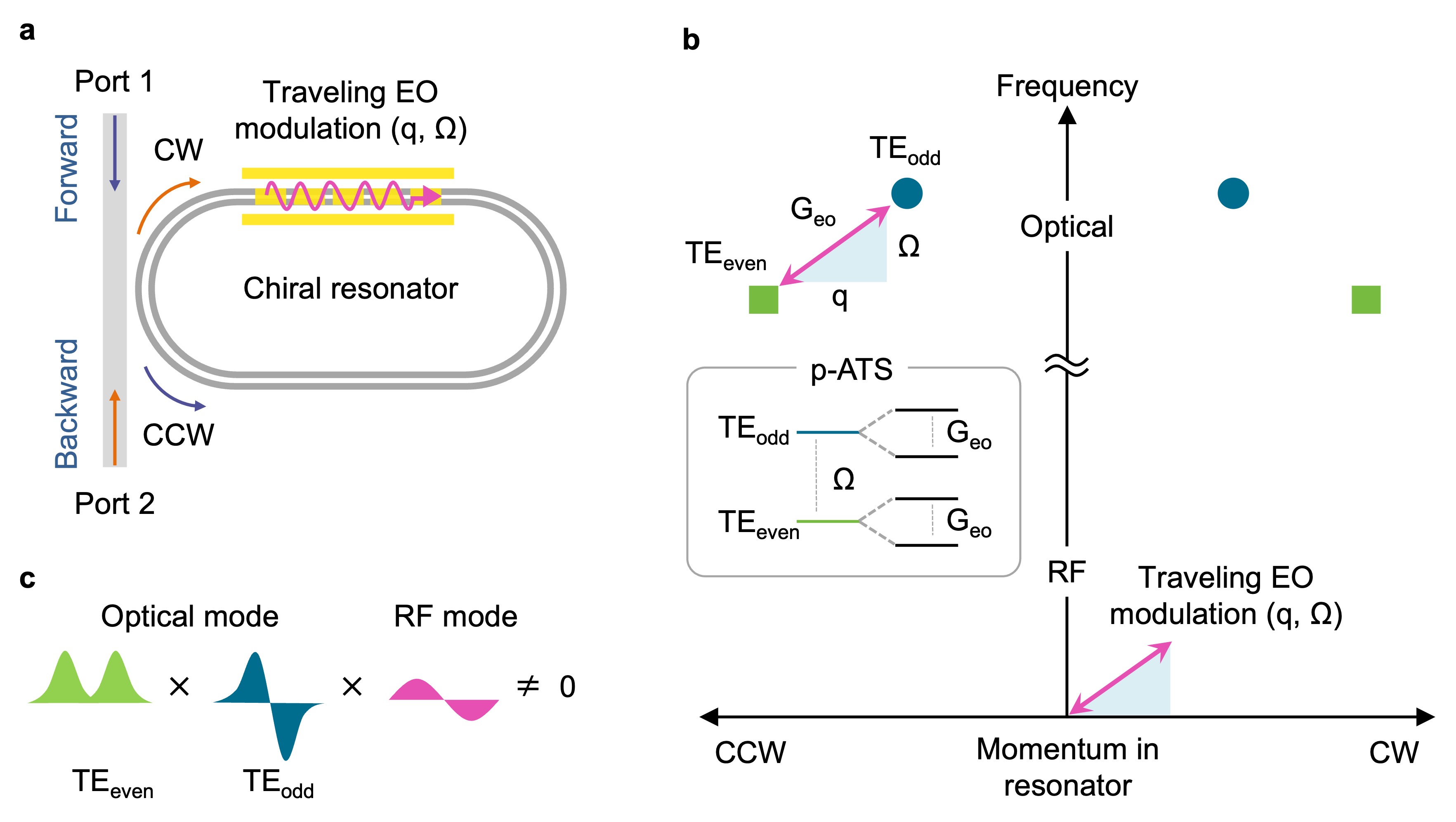}
    \centering
    \caption{
        \textbf{Isolator design and operational principle.}
        %
        \textbf{(a)} 
        The isolator is composed of a bus waveguide coupled to a chiral photonic resonator whose optical density of states is made strongly asymmetric using an electro-optic (EO) modulation. Light propagating in the forward direction along the waveguide experiences no interaction with the resonator, while simultaneously light propagating in the backward direction is resonantly absorbed. 
        \textbf{(b)} 
        This energy-momentum diagram shows the optical modes of the resonator configured to be separated in both frequency and momentum space, forming a two-level photonic atom. The EO stimulus generates coupling rate $G_{eo}$ for one direction of circulation, resulting in a strongly chiral DoS due to photonic Autler-Townes splitting (p-ATS).
        \textbf{(c)} 
        This illustration conveys that the RF stimulus must be designed with odd symmetry to help break the orthogonality between the optical modes of the resonator.
    }
    \label{fig1}
    \end{adjustwidth*}
\end{figure}

At the core of this isolator is a two-level ``photonic atom''~\cite{zhang_electronically_2019} whose optical density of states (DoS) exhibits large chiral asymmetry due to an induced photonic Autler-Townes splitting (p-ATS) effect \cite{peng_what_2014, kim_complete_2017, shi_nonreciprocal_2018, zhang_electronically_2019, sohn_electrically_2021}. This photonic atom is shunt coupled to a bus waveguide to produce isolation (Fig.~\ref{fig1}a). A straightforward approach to producing the photonic atom is with a whispering gallery or racetrack resonator that supports only two non-degenerate optical mode families with distinct dispersions. A unidirectional stimulus with non-zero propagating momentum is then applied to the resonator, ensuring that the phase-matching for inter-modal conversion is only satisfied for one direction of propagation (Fig.~\ref{fig1}b). When the stimulus-induced coupling rate $G_\textrm{stim} > \sqrt{\kappa_1 \kappa_2}$, where $\kappa_{1,2}$ are the loss rates for the individual optical modes, the system enters the strong coupling regime where the non-degenerate optical modes are hybridized and exhibit splitting, analogous to the splitting of electronic states in the conventional ATS phenomenon (a detailed coupled mode theory treatment can be found in the Supplementary Materials in {\cite{sohn_electrically_2021}}). 
Importantly, this mode splitting effect modifies the optical DoS only for the phase-matched direction \cite{sohn_electrically_2021, kim_complete_2017, shi_nonreciprocal_2018}. For the counter-propagating direction, the original optical DoS remains unmodified. 
The resonator is then used as a chiral absorber by critical shunt coupling to a waveguide so that, for some range of wavelengths, there is no interaction with light propagating in the forward direction along the waveguide, but simultaneously there is very strong extinction for light propagating in the backward direction. 
Importantly, all undesirable modulation sidebands are strongly suppressed as they are not phase-matched to any of the available optical states. Coupled mode theory for ATS type isolators~{\cite{sohn_electrically_2021}} predicts that forward insertion loss and the residual first-order sideband in the forward direction both diminish monotonically with increasing drive strength, reaching fully linear isolator behavior as $G_\textrm{stim} \rightarrow \infty$. %
    
The previous AO implementations of this concept used single input propagating acoustic wave {\cite{sohn_electrically_2021}} and 3-input acoustic standing wave {\cite{tian_magnetic-free_2021}} stimuli to accomplish the phase-matching. Here we aim to replace this with an EO stimulus to facilitate improved tunability and robust cladding. However, unlike acoustic phonons which possess substantial momentum \mbox{\cite{sohn_electrically_2021, kittlaus_electrically_2020}}, RF traveling waves have very high phase velocity and are unable to bridge the required momentum gaps at practical frequencies. To date, there has been no demonstration of \mbox{\it non-reciprocal} strong coupling in an integrated electro-optic device -- in either waveguide or resonator geometry -- which can be potentially transformative for non-reciprocal photonics technologies. %

\begin{figure}[hp]
    \begin{adjustwidth*}{-1in}{-1in}
    \hsize=\linewidth
    \includegraphics[width=1.3\columnwidth]{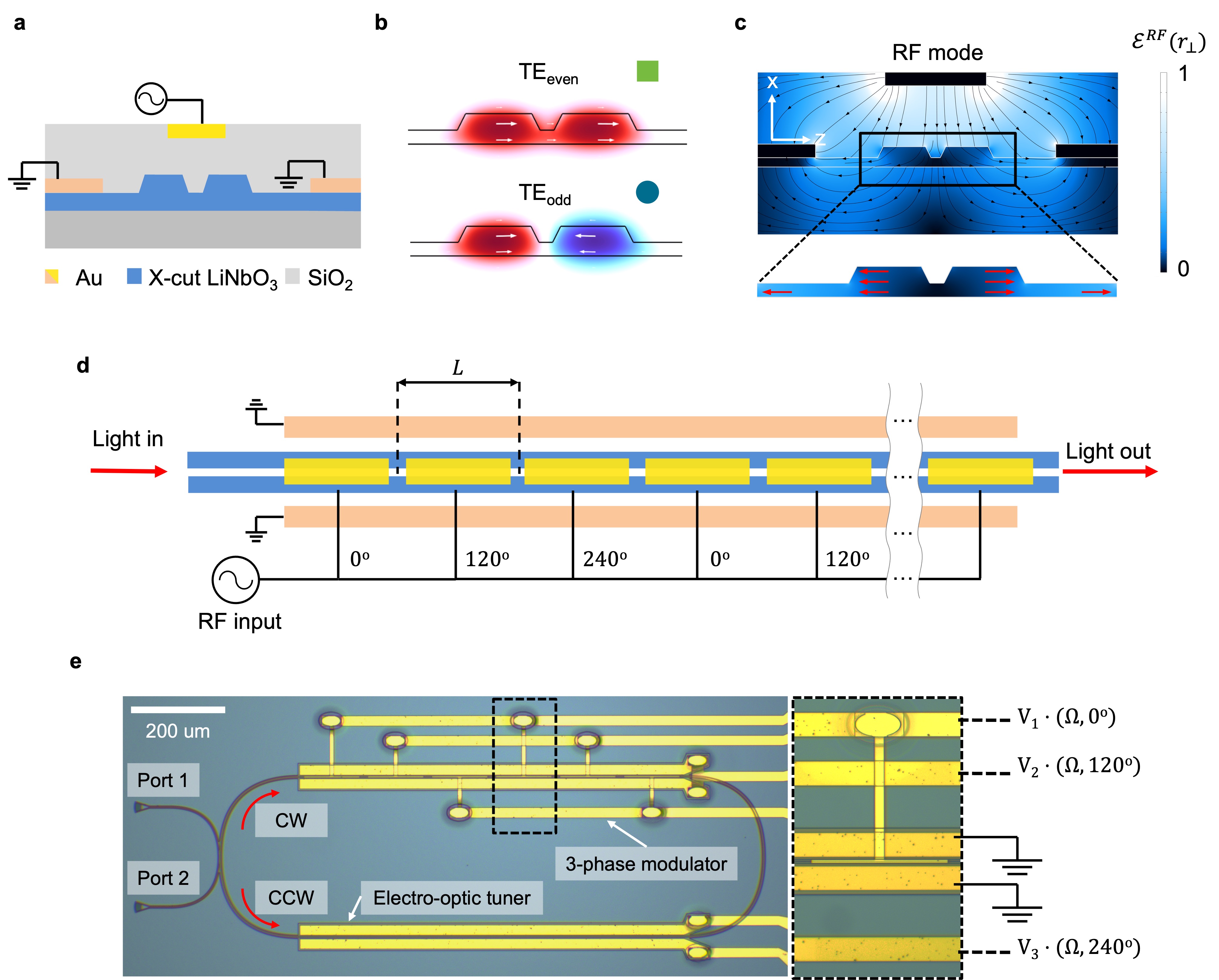}
    \centering
    \caption{
        \textbf{Design and implementation of the electro-optic isolator.}
        \textbf{(a)} 
        The cross-sectional schematic of the double-racetrack resonator shows the lithium niobate (LN) waveguide structure (dark blue), electrode configuration (yellow/orange), and oxide cladding (grey). 
        \textbf{(b)} 
        Finite element simulation of the TE\textsubscript{even} and TE\textsubscript{odd} optical modes supported by the resonator. 
        \textbf{(c)}
        Finite element simulation of the RF mode showing the transverse mirror symmetric electric-field profile.
        \textbf{(d)}
        A plan-view schematic of the 3-phase EO stimulus section is presented. The split-electrode structure is driven through three RF bus lines with $120^{\circ}$ relative phase shift with matched amplitudes at driving frequency $\Omega$. 
        \textbf{(e)}
        A top-view optical microscope image of the fabricated EO isolator. Upper section of the racetrack resonator shows the 3-phase EO modulator with the inset showing a zoomed-in view of a single electrode. Lower section of the resonator enables EO fine-tuning of the isolator.
    }
    \label{fig2}
    \end{adjustwidth*}
\end{figure}

Our solution is to invoke a periodic EO modulator that generates a synthetic stimulus momentum {\cite{orsel_electro-optic_2023}}. This approach also alleviates the phase-matching challenge across a wide spectrum by making the frequency a ``free choice'' variable that can be selected appropriately during operation.
The schematic for our specific modulator is shown later in Fig.~\ref{fig2}d, where a controlled relative phase shift $\Delta\phi = 120^{\circ}$ is applied between subsequent electrodes having pitch $L$. This configuration establishes a dominant spatial wavenumber of $q = \Delta\phi / L$ (harmonics are also generated but are rejected by the resonator). The stimulus momentum is therefore set by an initial lithographic choice of $L$, while the stimulus frequency $\Omega$ is freely selectable during operation. 
This synthetic momentum $q$ directly replaces the acoustic momentum from the AO designs \mbox{\cite{sohn_electrically_2021, tian_magnetic-free_2021}}.

The next major requirement is to ensure a fixed azimuthal momentum separation between the optical mode pairs over a wide spectral range. The previous AO designs~\cite{sohn_electrically_2021, tian_magnetic-free_2021} required precise dimension control of the resonator to obtain the required momentum separation between the modes, but this causes the device yield to be very sensitive to small fabrication variations. Moreover, since the optical dispersions are different, the phase matching in a selected device is only achievable between a limited set of optical mode pairs, which restricts dynamic tuning of the isolation wavelength. Accordingly, our second key innovation is to instead use a double-racetrack resonator \cite{xu_mutual-coupling_2022, zhang_broadband_2022}, composed of identical component waveguides having identical underlying dispersion (Fig.~\ref{fig2}a). This allows us to very easily control the momentum separation of the TE\textsubscript{even} and TE\textsubscript{odd} optical modes (Fig.~\ref{fig2}b) across a very wide spectrum, even though the frequency separation can vary slightly (as seen later in Fig.~\ref{fig5}). 
The combination of the above two design features -- the freedom to choose the stimulus frequency, and the very steady momentum separation -- enables the isolation effect to be coarsely tuned to optical mode pairs across a very wide spectral range.

\setlength{\abovedisplayskip}{3pt}
\setlength{\belowdisplayskip}{3pt}

Finally, the RF stimulus needs to be designed so that it breaks orthogonality (Fig.~\ref{fig1}c) between the TE\textsubscript{even} and TE\textsubscript{odd} modes via the $\chi^{(2)}$ nonlinearity. This is interpreted as a non-zero overlap integral between the optical modes and the RF stimulus, which is mathematically expressed as:
\begin{align}
\int \sum_{ijk} \mathcal{E}_i^\textrm{even}(r_\bot) \, \mathcal{E}_j^\textrm{odd}(r_\bot) \,\mathfrak{r}_{ijk} \, \mathcal{E}_k^\textrm{RF}(r_\bot) \, d{r_\bot} \neq 0 
\end{align}
where $\mathcal{E}^\textrm{even}({r_\bot})$, ${\mathcal{E}}^\textrm{odd}({r_\bot})$, and ${\mathcal{E}}^\textrm{RF}(r_\bot)$ are the transverse field distribution of optical and RF modes respectively (Fig.~\ref{fig2}b,c) and the subscripts $ijk$ refer to coordinate axes for the electro-optic tensor $\mathfrak{r}_{ijk}$ (the mathematical details are presented in \cite{orsel_electro-optic_2023}). For our specific device, the best RF electrode configuration requires a two-layer structure. The lower layer supports the uninterrupted ground electrodes while the upper layer is split into the multi-phase stimulus electrode (Fig.~\ref{fig2}a,d).

\vspace{12pt}
\section*{Results}

We fabricated this optical isolator (Fig.~\ref{fig2}e) using a 600 nm X-cut thin-film lithium niobate on insulator (TFLN) integrated photonics platform with an oxide cladding and two-layer gold electrodes to generate the required electro-optic stimuli. The fabrication process and key geometric features are described in Methods. Lithium niobate is well-suited for electro-optics due to its large EO coefficient ($r_{33} \approx 30$ pm/V) and wide bandgap, offering a broad transparency range of 250 to 5300 nm with propagation losses below 0.1 dB/cm \cite{weis_lithium_1985, desiatov_ultra-low-loss_2019}. The resonator design targets an optical momentum gap of 13000 -- 15000 rad/m corresponding to an azimuthal order difference of $\Delta M_\textrm{optical} = 7$. Accordingly, two periods of the 3-phase RF stimulus electrodes were patterned on one side of the resonator, and their pitch was selected at $L=150$ {\textmu}m, resulting in a total interaction length of 900 {\textmu}m (stimulus wavelength is then 450 {\textmu}m, which matches $M_\textrm{stimulus} = 7$ around the racetrack). The stimulus electrode cross-section (Fig.~\ref{fig2}a) was designed to generate a mirror-symmetric transverse electric field profile along the LN Z-axis (Fig.~\ref{fig2}c) to efficiently harness the $r_{33}$ coefficient for modulation. We additionally fabricated a set of EO tuning electrodes on the opposite side of the resonator.

\begin{figure}[ph!]
    \begin{adjustwidth*}{-1in}{-1in}
    \hsize=\linewidth
    \includegraphics[width=1.15\columnwidth]{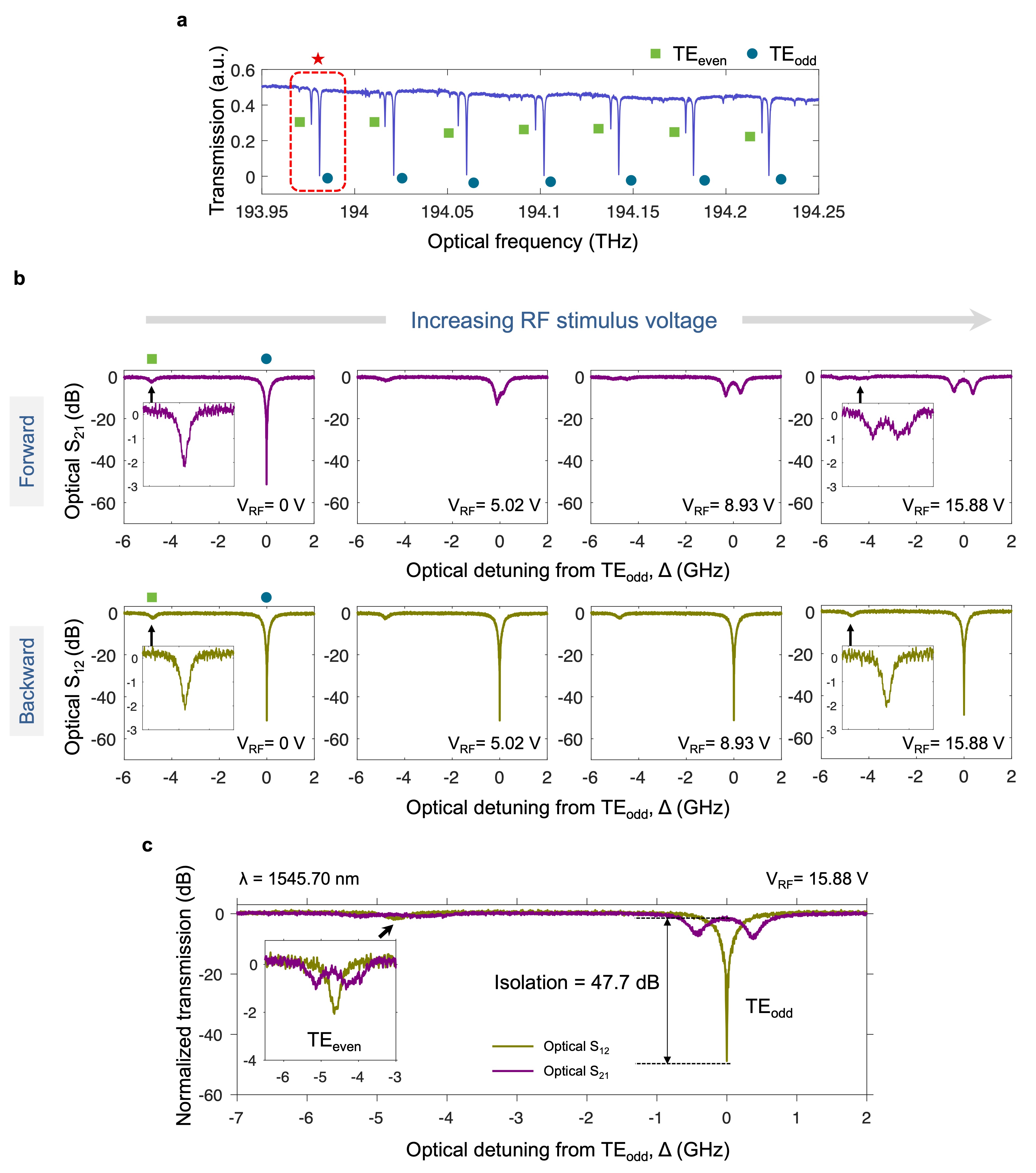}
    \centering
    \caption{
        \textbf{Experimental demonstration and characterization of isolation.}
        \textbf{(a)} Transmission measured through the waveguide shows the two optical mode families hosted by the resonator. We chose the optical mode pair marked by $\color{red}\bm{\star}$ located near 193.98 THz (1545.5 nm) to conduct our experiments. 
        \textbf{(b)} 
        With RF stimulus set to $\Omega = 4.8$ GHz, which matches the modal frequency gap, we measured evolution of the through-waveguide transmission for the $\color{red}\bm{\star}$ mode pair.       
        As the applied RF voltage (V\textsubscript{RF}) is increased, the p-ATS phenomenon appears for forward light propagation direction while the backward direction remains strongly absorbing. 
        \textbf{(c)} We present a detailed view of the superimposed forward and backward measurements for the 15.88 V case, corresponding to a measured $G_{eo}$ of approximately 0.89 GHz.
    }
    \label{fig3}
    \end{adjustwidth*}
\end{figure}

\begin{figure}[ph!]
    \begin{adjustwidth*}{-1in}{-1in}
    \hsize=\linewidth
    \vspace{-2cm}   
    \includegraphics[width=0.92\columnwidth]{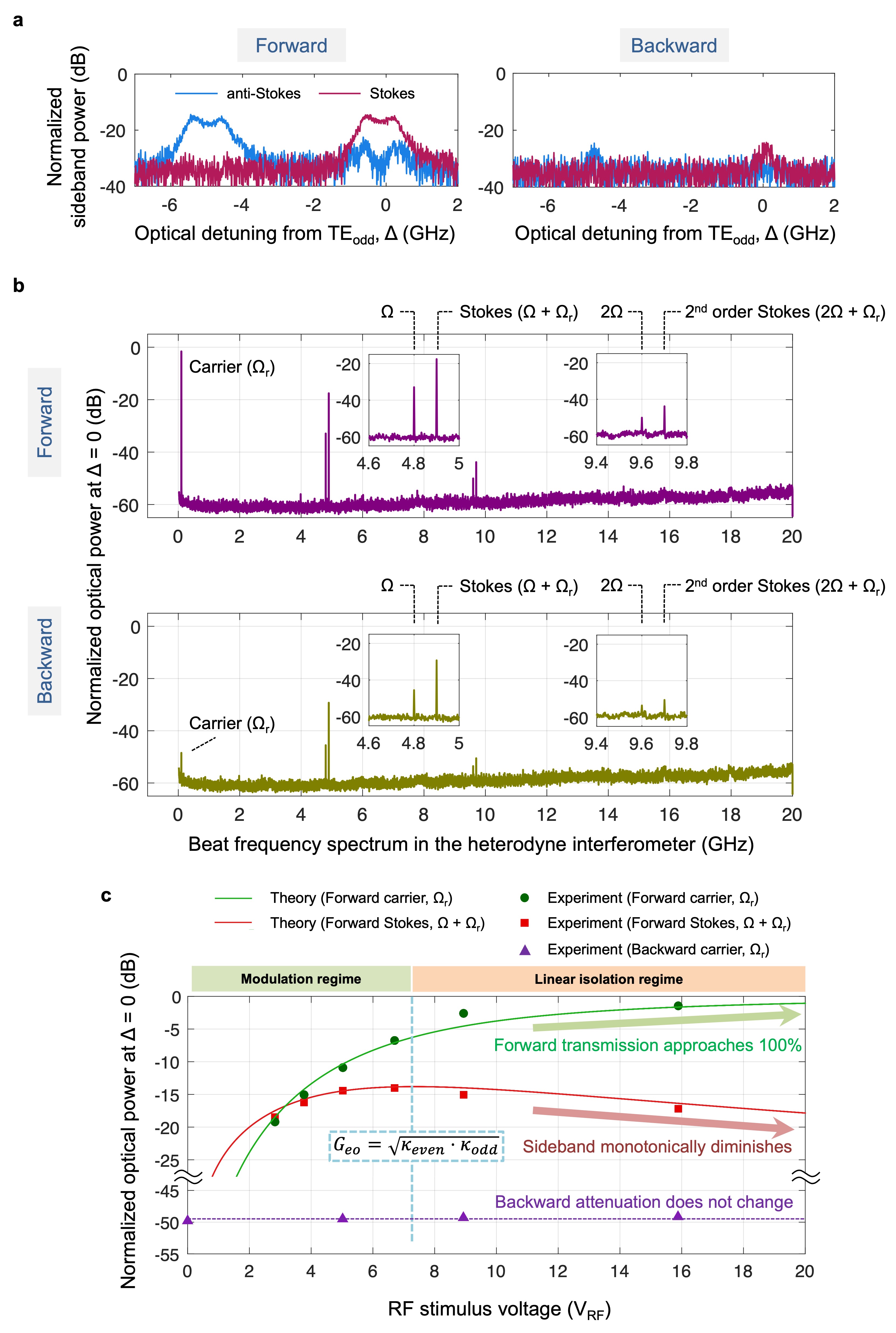}
    \centering
    \caption{
        \textbf{Sideband analysis during isolator operation.}
        \textbf{(a)} 
        The measured first-order sidebands as a function of optical detuning, normalized to the input optical carrier, for the case shown in Fig.~\ref{fig3}c.   
        \textbf{(b)}  
        These beat spectra obtained from the heterodyne detection system reveal the carrier and its sidebands (see Supplementary Data Fig.~\ref{Heterodyne}), normalized to the input optical carrier. Measurements are performed with a $\Omega_r=100$ MHz frequency shift applied to the reference arm of the interferometer; thus $\Omega_r$ measures the carrier, $\Omega\pm\Omega_r$ correspond to the first-order Stokes and anti-Stokes sideband respectively, and $2\Omega\pm\Omega_r$ correspond to the second-order Stokes and anti-Stokes sideband respectively. The anti-Stokes sidebands are notably absent, and beat notes at $\Omega, 2\Omega$ correspond to parasitics in the measurement system.
        \textbf{(c)}
        We plot the normalized carrier transmission and the Stokes sideband power in forward direction, and backward carrier transmission, all normalized to input optical carrier, as a function of RF voltage at $\Delta$ = 0. The solid line is a prediction from coupled mode theory \cite{sohn_time-reversal_2018, sohn_electrically_2021} using the experimentally measured device parameters.
    }
    \label{fig4}
    \end{adjustwidth*}
\end{figure}

\vspace{12pt}

\noindent
\textbf{Demonstration of electro-optic isolation --}
Fig.~\ref{fig3} presents an example experiment on the isolator. First, we measured optical transmission through the waveguide in the telecom band, around 1550 nm, which reveals two distinct optical mode families of the resonator (Fig.~\ref{fig3}a) with separation of a few GHz. The modes can be identified as TE\textsubscript{even} and TE\textsubscript{odd} based on both coupling to the waveguide and their dispersive effect supported through finite element simulations. Next, we experimentally measure the through-waveguide transmission using a heterodyne detection system (see Methods) which enables resolved measurements on the carrier transmission, as well as any sideband signals that are generated. In this experiment we chose an optical mode pair near 193.98~THz (1545.50~nm) as identified in Fig.~\ref{fig3}a with frequency gap of approximately 4.8~GHz and individual loss rates $\kappa_\textrm{even}=0.28$ GHz and $\kappa_\textrm{odd}=$ 0.53~GHz. The TE$\textsubscript{odd}$ mode is critically coupled and provides large extinction for the waveguide transmission, while the TE$\textsubscript{even}$ mode is more inward located and is therefore undercoupled. We accordingly apply a 3-phase RF stimulus at $\Omega = 4.8$~GHz to the electrodes. We also note that this RF drive frequency does not have to be precisely matched to the frequency gap and the isolator continues to operate as long as the system of modes remains in the strong coupling regime {\cite{sohn_electrically_2021}}. Fig.~\ref{fig3}b presents the evolution of the optical transmission through waveguide as a function of RF voltage (i.e. sinusoidal amplitude of RF stimulus, V\textsubscript{RF}) for both forward and backward directions. Here, we arbitrarily set the TE\textsubscript{odd} mode as the zero detuning reference point ($\Delta=0$) for the optical frequency axis. 
In the absence of any applied RF, the transmission function remains identical in both directions. However, with increasing RF voltage, the p-ATS modal hybridization appears for one direction of propagation only. In Fig.~\ref{fig3}c we present a superimposed result at the highest applied RF amplitude of 15.88 V, reaching a peak isolation contrast of 47.7 dB with a forward insertion loss of 1.45 dB. This implies IC/IL ratio of $\approx 32.9$ dB/dB which makes the performance comparable to commercial magneto-optic isolators. 
By fitting to coupled mode theory \cite{sohn_electrically_2021}, we determine that this driving condition corresponds to an EO coupling rate $G_{eo} \approx 0.89$ GHz. We measure a 10 dB IC bandwidth of $\approx 150$ MHz and 20 dB IC bandwidth of $\approx 40$ MHz. We tested this device with an estimated in-waveguide optical power up to 5 mW.

While the EO stimulus on this device has an overall traveling wave characteristic, the propagating momentum is generated only synthetically. We confirm this using a RF S\textsubscript{11} reflection measurement at the stimulus electrodes (Supplementary Data Figure~\ref{S11}) which confirms their uniformly high reflection and open circuit (capacitive) behavior \cite{orsel_electro-optic_2023}. The residual loss arises from propagation in the measurement cables. For the specific experiment in Fig.~\ref{fig3}, the bench-top signal generator power was set at 28 dBm (630 mW). 
Since the load at the device is a capacitor (i.e. open circuit), and since the source, cable, and RF probe impedances are all 50 ohms, this corresponds to a sinusoidal voltage on each electrode having amplitude
    $V_\textrm{RF} = 2 \times \sqrt{2 \cdot 50 \cdot P_{drive}} = 15.88$ V
\mbox{\cite{orsel_electro-optic_2023}}.
The RF S\textsubscript{11} measurement at 4.8 GHz informs us of a fundamental upper limit of only 12.9 mW (11.09 dBm) power consumed by the device. 
    With improved RF design (e.g. impedance matching circuits if using external signal generators) or a custom RF voltage generator on-chip, the power consumption could be reduced significantly.
Thus, essentially all of the signal generator voltage (15.88 V) appears directly on the stimulus electrodes due to the open circuit.

\vspace{12pt}

\noindent
\textbf{Sideband analysis --}
During experiments we also monitored the anti-Stokes and Stokes sidebands since they are a concern in non-reciprocal devices that use strong spatio-temporal modulation~\cite{doerr_optical_2011,doerr_silicon_2014,yu_integrated_2023, gao_thin-film_2024}.
Fig.~\ref{fig4}a presents measurements of the first-order sidebands as a function of optical detuning $\Delta$, under the operational conditions of Fig.~\ref{fig3}c. 
    Naturally, the sideband generation in the forward (phase-matched) direction is shaped by the optical DoS as modified by the mode splitting. 
In the limit of very strong coupling, the hybrid states will be fully resolved \cite{kim_complete_2017}, and at $\Delta=0$ the optical DoS will reach zero causing the forward direction first-order sideband to be eliminated.
In the backward (not phase-matched) direction the sidebands are significantly smaller.    

Fig.~\ref{fig4}b analyzes the higher-order sidebands in the optical spectrum using the heterodyne interferometer (see Methods and Supplementary Data Figure~\ref{Heterodyne}) with the input laser centered on the TE\textsubscript{odd} mode ($\Delta=0$) and reference arm frequency shift of $\Omega_r = 100$ MHz. We expect that only first-order Stokes scattering will be notable since only scattering to TE\textsubscript{even} can take place. 
The interferometer spectrum shows the carrier (at $\Omega_r$), first Stokes (at $\Omega+\Omega_r$), and a much smaller second-order Stokes (at $2\Omega+\Omega_r$) signal. We normalize the forward and backward measurements relative to the carrier transmission in the forward direction. These measurements confirm that only the first Stokes sideband is of significance at approximately -17.8 dB relative to the carrier. The first anti-Stokes ($\Omega-\Omega_r$) and second anti-Stokes ($2\Omega-\Omega_r$) are absent, or below the noise floor, as there are no optical states available within the resonator.

These sideband generation behaviors are readily predictable from coupled mode theory \cite{sohn_electrically_2021, kim_complete_2017} and two regimes of operation have been identified. When the applied stimulus $G_{eo}$ is small, the device acts as a non-reciprocal frequency shifting modulator \cite{sohn_time-reversal_2018}. However, once $G_{eo}>\sqrt{\kappa_\textrm{even}\cdot\kappa_\textrm{odd}}$ the device enters the strong coupling regime where linear isolation becomes available. With increasing drive level the hybrid states get resolved further and the ideality of the device monotonically improves as sideband generation reduces monotonically. This behavior and the close match between experiments and theory is confirmed in Fig.~\ref{fig4}c. 
Importantly, if the auxiliary mode (here the \mbox{TE\textsubscript{even}} mode) is fully decoupled from the bus waveguide, any residual scattered Stokes sideband can be completely eliminated. %

\begin{figure}[htp]
    \begin{adjustwidth*}{-1in}{-1in}
    \hsize=\linewidth
    \includegraphics[width=1.1\columnwidth]{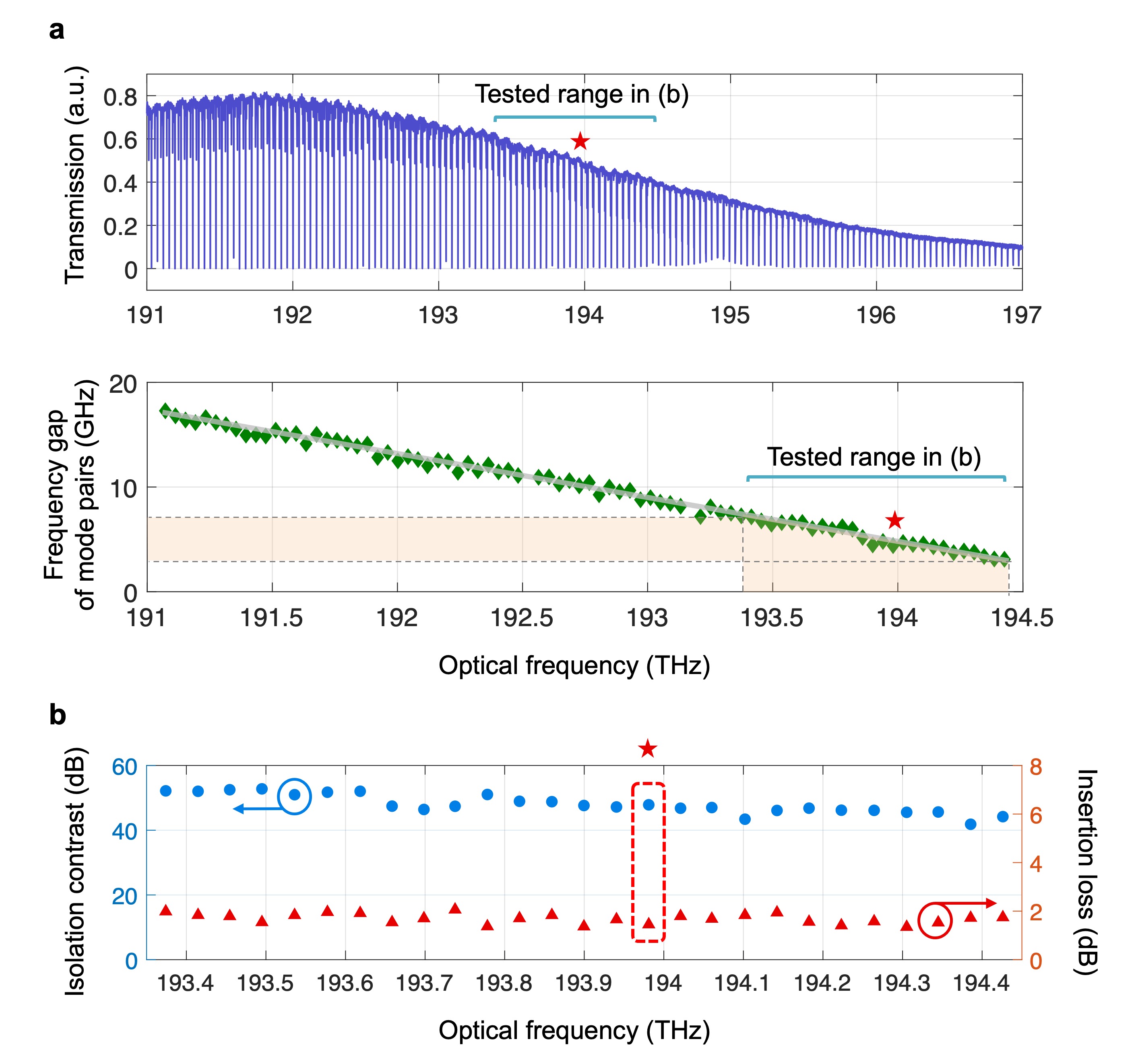}
    \centering
    \caption{
        \textbf{Frequency tuning of the isolator band.}
        \textbf{(a)} 
        Waveguide transmission measured over 191--197 THz (1520--1570 nm). The background transmission is shaped by the optical grating couplers. 
        The $\color{red}\bm{\star}$ indicates the mode pair examined in Figs.~\ref{fig3} and \ref{fig4}.
        We additionally measure the frequency gap of all the $\Delta M_\textrm{optical} = 7$ mode pairs over the range 191--194.5 THz (1542--1570 nm), confirming a linear relative disperion \mbox{$\sim$4.1 GHz/THz}. 
        \textbf{(b)} 
        The isolator is tested across a 1 THz band (8 nm wavelength range between 1542--1550 nm), with isolation contrast and insertion loss measured for 27 distinct mode pairs. The RF stimulus frequency $\Omega$ is set individually in each experiment to match the frequency gap for the mode pair, while V\textsubscript{RF} is fixed to 15.88 V. All measurements maintain $41-52$ dB contrast across this spectral range and show insertion loss $1.4-2.0$ dB. 
    }
    \label{fig5}
    \end{adjustwidth*}
\end{figure}

\vspace{12pt}

\noindent
\textbf{Tunability across a multi-THz spectrum --}
A wide transmission spectrum of the device, at the limits of our 1520--1570 nm tunable laser, is presented in Fig.~{\ref{fig5}a}. The frequency gap between modes of azimuthal order $\Delta M_\textrm{optical}=7$ 
evolves with a linear trend of $\sim$4.1 GHz/THz across the spectrum due to dispersion.
The shape of the background transmission is defined by the optical grating couplers and can be flattened if we instead used edge couplers.
We tested the tuning capability of the isolator over the range \mbox{193.35--194.45 THz} ($\sim$8 nm from 1542--1550 nm) which was only limited by the frequency constraint of our laboratory's RF signal generators. 
For all mode pairs in this range, we applied the 3-phase RF stimulus matching the frequency gap $\Omega$ with $V_\textrm{RF} = 15.88$ V. This is possible since, as mentioned earlier, the frequency is a completely free variable in this EO design. 
Fig.~\ref{fig5}b presents the measured isolation contrast and insertion loss. 
We confirm that the device has consistent optical isolation performance with $41-52$ dB contrast and $1.4-2.0$ dB insertion loss for all mode pairs across this spectral range. Thus, each mode pair indicates a coarse tuning choice for the isolator.

This coarse tuning capability is not constrained by any fundamental limit, and the isolation can extend far beyond the demonstrated span as long as an RF source is available to bridge the inter-modal frequency gap.
The spacing for $\Delta M_\textrm{optical}=7$ mode pairs near 191 THz is only $\sim$16 GHz (Fig.~\ref{fig5}a). 
The mode spacing approaches zero near 195 THz, which will be a detriment to isolator directionality, thus our device should have a minimum operational range of 191--194.5 THz. 
At higher optical frequencies $>$195 THz, the modes are again resolved but their spacing is negative which simply means that the isolator will operate in a flipped mode~\mbox{\cite{sohn_time-reversal_2018}}.
Assuming a continuing linear trend for the dispersions, the mode spacing should reach 60 GHz around at 180 THz optical frequency, which is readily accessible using microwave signal generators, and we can therefore speculate on a potential operational range of 180--194.5 THz (1542--1665 nm). Of course, it is additionally important to modify the device to a broadband optical edge coupling setup to circumvent the bandwidth limitations of the grating couplers which were used in our experiments.

Finally, we explore the possibility of fine-tuning to fill the spectral gaps between the coarse tuning points. We measure that a dc voltage on the co-integrated EO tuning electrodes (see Supplementary Data Fig.~\ref{EOtuning}) linearly shifts both modes of a pair together. 
For the device tested the tuning capability was measured at 0.20 GHz/V for the TE\textsubscript{odd} mode and 0.21 GHz/V for the TE\textsubscript{even} mode, with the difference arising due to the distinction of modal group index. The small monotonic change in the frequency gap can be tracked while tuning the device to appropriately adjust the stimulus frequency $\Omega$. Since the free spectral range between optical mode pairs in this device is around $40$ GHz, an application of $\pm$ 100 V\textsubscript{DC} can in principle electro-optically tune the modes to completely cover the gaps. We leave further improvements on the tuning structure to future efforts, since the resonator free spectral range can be reduced by simple redesign, and we additionally note that the combination of EO tuning and stress-based piezoelectric tuning~\cite{tian_piezoelectric_2024} could provide enhanced tuning capability.
The multi-THz tuning capability, in conjunction with state-of-the-art IC/IL figure of merit, implies unique isolation coverage that far surpasses any previous on-chip isolator.

\section*{Discussion}

This EO isolator is the first integrated photonic device to demonstrate \textit{non-reciprocal} strong coupling with electro-optic stimulus. 
Since isolators can be cascaded, a highly valued figure of merit is the contrast that can be obtained per dB of insertion loss (ratio IC/IL in dB/dB) for the optical carrier signal. Commercial off-chip magneto-optic isolators typically exhibit IC/IL $>30$ dB/dB, a level that to date only two reported non-magnetic chip-scale isolators have reached~\mbox{\cite{sohn_electrically_2021, yu_integrated_2023}} (see Supplementary Data Table {\ref{tab:FoM_table}}).
The device we demonstrate here also reaches this figure-of-merit regime while using nearly 20x smaller linear footprint (900~{\textmu}m here, vs 1.75~cm) and generating far fewer sidebands than the best comparable EO isolator {\cite{yu_integrated_2023}}. %
The efficiency of our device can be improved further by incorporating more periods of the EO stimulus electrodes around the resonator resulting in a larger interaction length. The EO tuning could then be performed using these same stimulus electrodes but with two different dc ground offset voltages.
As with the AO chiral shunt isolator, our EO design has an instantaneous isolation bandwidth that is fundamentally set by the optical modes, though that could be improved somewhat by lowering the optical Q-factors while simultaneously increasing $G_{eo}$. Specifically, the maximum isolation bandwidth for our device is limited to the linewidth of the TE$\textsubscript{odd}$ mode, even for infinite $G\textsubscript{eo}$. 
We also showed that very wide tuning capability is enabled thanks to dual innovations on a double-racetrack resonator design and a synthetic momentum approach, both of which together relax the phase-matching constraints.
Thus, future modulation efficiency enhancements with novel materials like BaTiO\textsubscript{3}~\cite{dong_monolithic_2023,demkov_ferroelectric_2024,mohl_bidirectional_2025}, which can now reach the same required Q-factors with much higher Pockels coefficient \cite{kim_low_2025}, could improve ideality and performance metrics even further. Another approach to reduce the residual sideband is to fully decouple the auxiliary TE\textsubscript{even} mode from the waveguide by means of smart optical design.
Ultimately, these narrowband but widely tunable non-magnetic isolators can have a big positive impact for single-frequency protection, e.g. for ultrastable laser sources~\cite{poulton_coherent_2017,spencer_optical-frequency_2018,lucas_ultralow-noise_2020}, atomic references~\cite{knappe_microfabricated_2004,hummon_photonic_2018,newman_architecture_2019}, and optical metrology~\cite{delhaye_optical_2007}.

\vspace{2cm}

\section*{Methods}
\label{sec:Method}

\textbf{Device fabrication --} The starting substrate was X-cut thin film lithium niobate on 2 {\textmu}m silicon dioxide on a silicon handle wafer. Optical components were patterned using a 150 keV electron beam lithography (EBL, Elionix ELS-G150) system and the thin film lithium niobate was etched down by 300 nm (approximately 60 degree sidewalls) via argon inductively coupled plasma reactive ion etch (ICP-RIE). 
The double-racetrack resonator was designed using identical concentric waveguides, each having width 1.2 {\textmu}m and gap of 650 nm between them.
Lower electrodes were defined by EBL using 200 nm gold deposited through e-beam evaporation and lift-off process, and placed 3 {\textmu}m away from the edge of the double racetrack resonator. A 2 {\textmu}m silicon dioxide (SiO\textsubscript{2}) cladding layer was deposited using plasma enhanced chemical vapor deposition (PECVD). Metal via holes were created in the cladding using buffered oxide etching. Top electrodes were finally patterned by depositing an additional 400 nm gold layer via e-beam evaporation and lift-off process.

\vspace{12pt}

\noindent
\textbf{Heterodyne measurement system -- } We measure the carrier and sidebands using a heterodyne interferometer (Supplementary Data Fig.~\ref{Heterodyne}) which was described in \cite{sohn_electrically_2021}. A tunable external cavity diode laser (New Focus TLB-6712-P), operating within the 1520–1570 nm range and featuring a sub-50 kHz linewidth, was utilized as the light source. An acousto-optic frequency shifter (Brimrose TEF-200-780-2FP) was employed to generate a reference signal for frequency-resolved characterization of the carrier and sidebands, which were analyzed using a high-speed photodetector (Thorlabs RXM25DF). Probe direction was managed via an off-chip optical switch (Thorlabs OSW22-780E). The optical beat signal was generated and analyzed using a 4-port vector network analyzer (Agilent ENA E5080A) and a real-time spectrum analyzer (Tektronix RSA5100B). RF signals were directly supplied by a multi-output synchronized microwave signal generator (Holzworth 9004).

\end{bibunit}

%
\section*{Acknowledgments}

This work was sponsored by the Defense Advanced Research Projects Agency (DARPA) under Cooperative Agreement D24AC00003, the Air Force Research Laboratory (AFRL) / US Space Force (USSF) grant FA9453-20-2-0001, the US Office of Naval Research (ONR) Multi-University Research Initiative grant N00014-20-1-2325, and the Army Research Office (ARO) grant W911NF-23-1-0219. The views and conclusions contained herein are those of the authors and should not be interpreted as necessarily representing the official policies or endorsements, either expressed or implied, of DARPA, AFRL, ONR, ARO, or the US Government.
The authors would like to acknowledge invaluable support from Dr. Brian Kasch at the Air Force Research Laboratory, Space Vehicles Directorate, and Dr. Edmond Chow and Amr O. Ghoname at the Holonyak Micro \& Nanotechnology Lab (HMNTL) at University of Illinois at Urbana-Champaign.

\section*{Data availability}

The data that support the findings of this study are available from the corresponding authors upon reasonable request.

\section*{Author contributions}

G.I.K. and G.B. jointly conceived the concept, G.I.K. and V.W. performed the theoretical analysis. G.I.K., O.E.Ö., and J.Y. contributed to the device fabrication and the experimental measurements. G.I.K. and V.W. analyzed the experimental data. All authors contributed to writing the paper. G.B. supervised all aspects of this project.

\FloatBarrier

\newpage

\newcommand{\beginExtFigures}{%
        \setcounter{table}{0}
        \renewcommand{\tablename}{Supplementary Data Table}
        \renewcommand{\thetable}{\arabic{table}}%
        \setcounter{figure}{0}
        \renewcommand{\figurename}{Supplementary Data Figure}        
        \renewcommand{\thefigure}{\arabic{figure}}%
}
\beginExtFigures

\begin{figure}[h]
    \begin{adjustwidth*}{-1in}{-1in}
    \hsize=\linewidth
    \includegraphics[width=1.2\columnwidth]{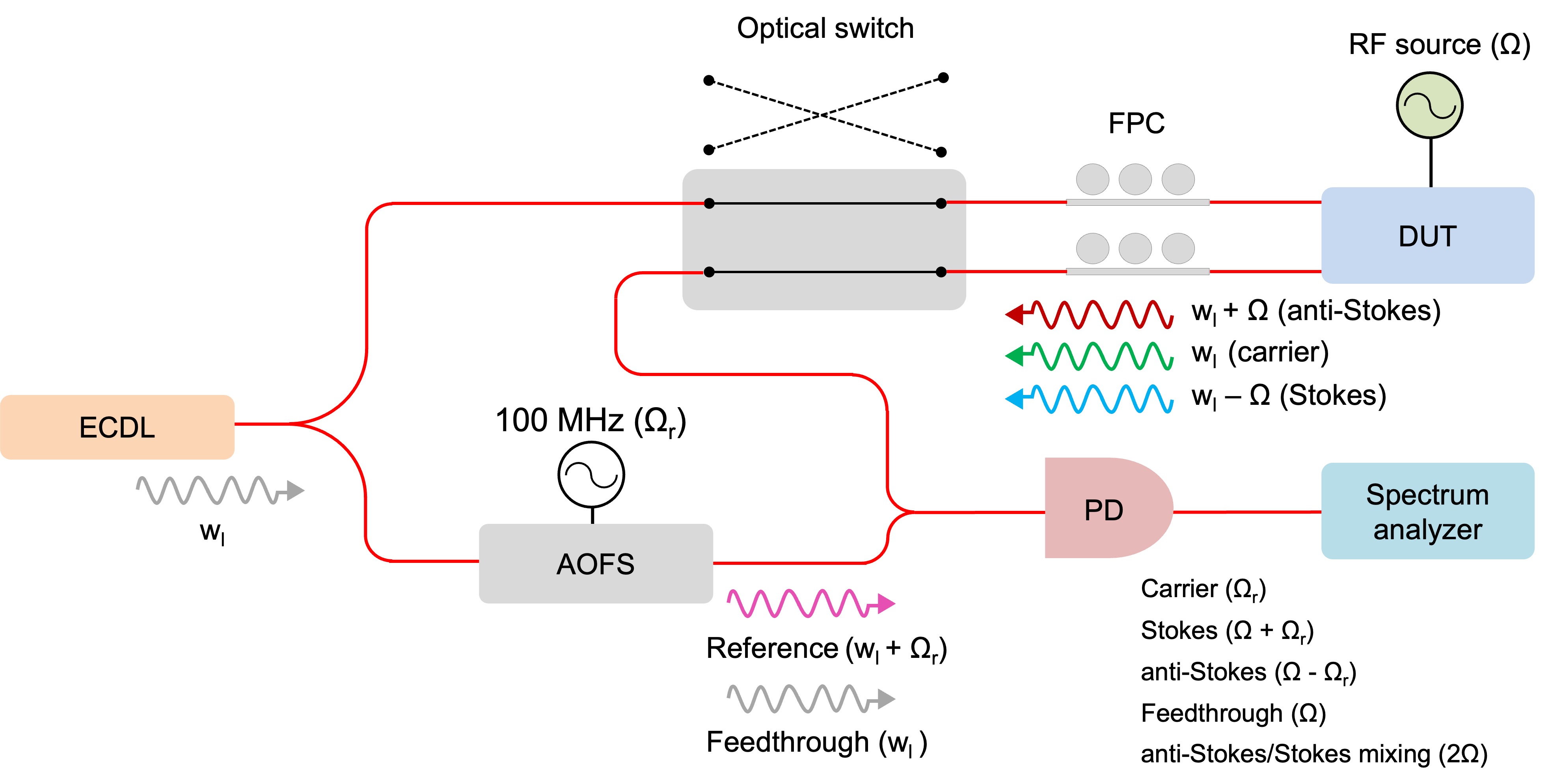}
    \centering
    \caption{
        Heterodyne detection system used for measuring the carrier and sidebands. ECDL = External cavity diode laser. DUT = Device under test. FPC = Fiber polarization controller. AOFS = Acousto-Optic Frequency Shifter. PD = Photodetector.}
    \label{Heterodyne}
    \end{adjustwidth*}
\end{figure}

\begin{figure}[h]
    \begin{adjustwidth*}{-1in}{-1in}
    \hsize=\linewidth
    \includegraphics[width=0.5\columnwidth]{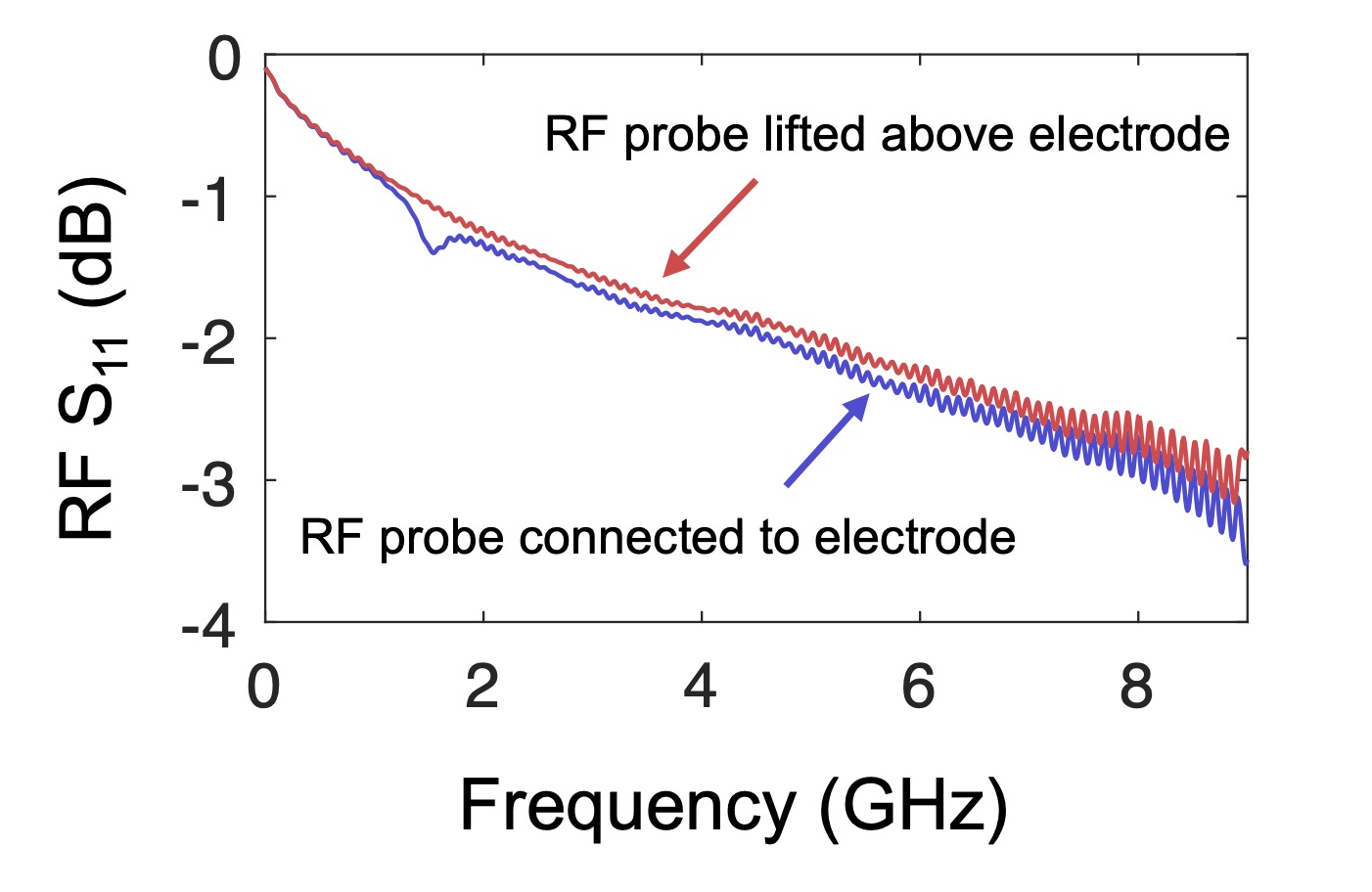}
    \centering
    \caption{
        Measurement of the RF reflection coefficient (S\textsubscript{11}) at the electrodes. The slight dip observed around 1.5 GHz arises from resonance effects in the cables, attributed to back-reflections from the electrode.
    }
    \label{S11}
    \end{adjustwidth*}
\end{figure}

\FloatBarrier
\newpage

\begin{figure}[h]
    \begin{adjustwidth*}{-1in}{-1in}
    \hsize=\linewidth
    \includegraphics[width=0.85\columnwidth]{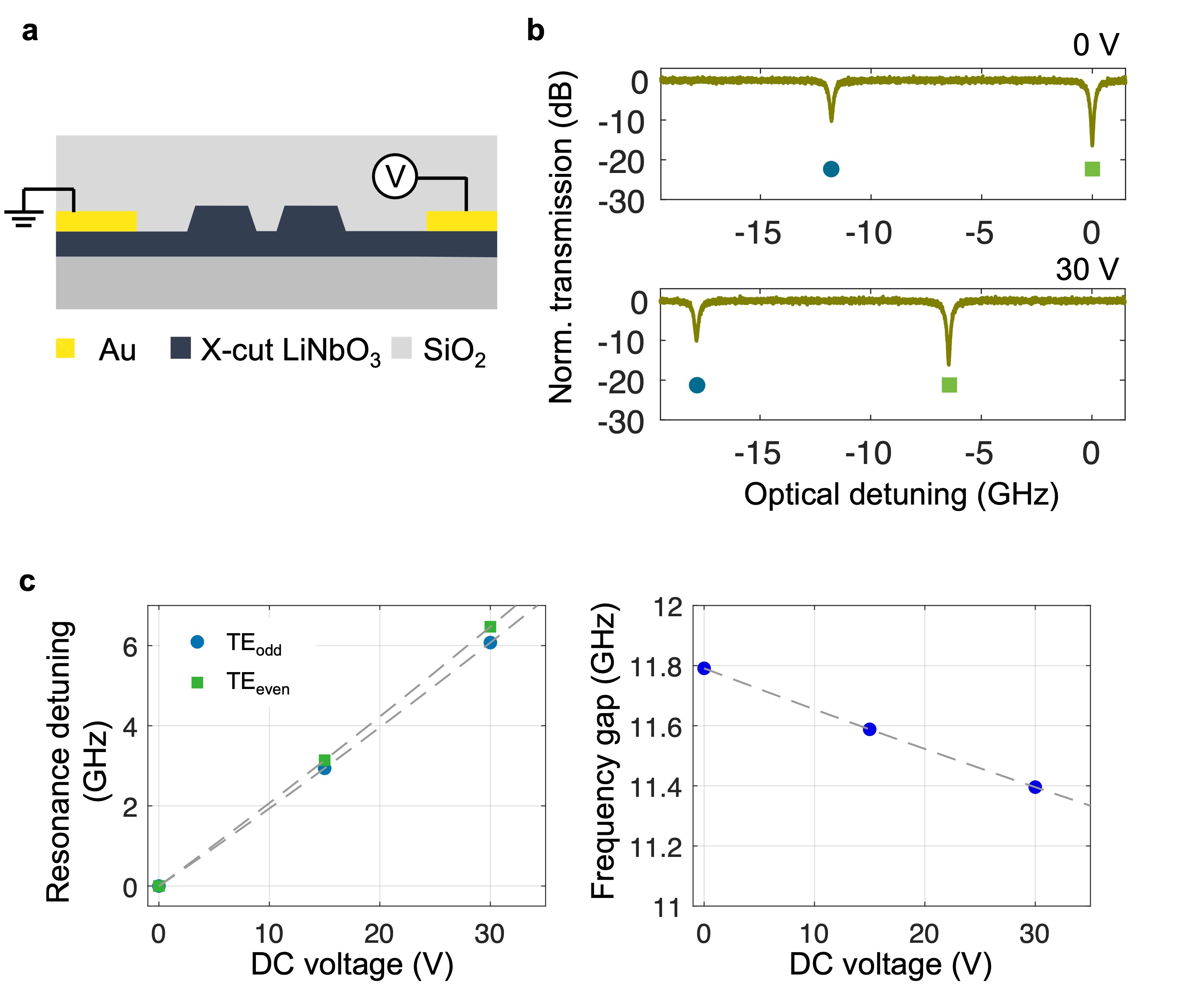}
    \centering
    \caption{
        \textbf{(a)} 
        The cross-sectional schematic of the EO tuner on the ring resonator (shown in Fig.~\ref{fig2}e).
        \textbf{(b)} 
        Example transmission measurements through the waveguide, on a device with mode pair spacing around 11.5 GHz, showing how the resonator's even and odd modes shift with applied dc voltage.
        \textbf{(c)} 
        Consolidated tracking of the individual mode detuning and evolution of the frequency gap as a function of applied dc voltage.
    }
    \label{EOtuning}
    \end{adjustwidth*}
\end{figure}

\FloatBarrier

\vspace{1cm}

\newcommand{\NA}[0]{\cellcolor{black!15}N/A}
\newcommand{\AO}[0]{\cellcolor{green!15}AO}
\newcommand{\EO}[0]{\cellcolor{blue!15}EO}
\newcommand{\MO}[0]{\cellcolor{pink!15}MO}
\newcommand{\ON}[0]{\cellcolor{yellow!15}ON}
\newcommand{\blue}[0]{\cellcolor{blue!15}}
\newcommand{\green}[0]{\cellcolor{green!15}}

	\begin{centering}
		\begin{table}[h]
		\small

			\caption{Comparison between published EO and electrically driven AO isolators in integrated photonics. Results with IC/IL $>30$ dB/dB, i.e. competitive with commercial isolators, are marked in bold.} 

			\begin{adjustwidth}{-1.1in}{-0.5in}
			\begin{tabular}{| >{\raggedleft\arraybackslash}m{3.0cm} | >{\raggedleft\arraybackslash}m{0.8cm} | >{\raggedleft\arraybackslash}m{1.6cm} | >{\centering\arraybackslash}m{1.5cm} | >{\raggedleft\arraybackslash}m{1.9cm} | >{\raggedleft\arraybackslash}m{4.2cm} | >{\raggedleft\arraybackslash}m{1.8cm} |}
				
				\hline
				Author & Year & Device & Technique & Isolation contrast (IC) & Insertion loss (IL) & IC/IL (dB/dB)  \\ \hline \hline
				
				This work & - & Resonator & \EO & 47.7 dB & 1.45 dB & \textbf{32.9} \\ \hline
				\hline

                Yu et. al.~\cite{yu_integrated_2023}& {2023} & Waveguide & \EO & 48 dB & 1.4 dB (w/ filter)  & \textbf{34.28} \\
                & & & \blue &  &  0.5 dB (w/o necessary filter) & \textbf{96} \\ \hline
                Gao et. al.~\cite{gao_thin-film_2024}& 2023 & Waveguide & \EO & 39.5 dB & 2.6 dB (w/o filter) & 15.19 \\ \hline
                Dostart et. al.~\cite{dostart_optical_2021}& 2021 & Resonator & \EO & 13.1 dB & 18.1 dB & 0.18 \\  \hline
                Doerr et. al.~\cite{doerr_silicon_2014}& 2014 &  Waveguide & \EO & 6 dB & 4 dB & 0.66 \\ \hline
                Tzuang et. al.~\cite{tzuang_non-reciprocal_2014}& 2014 & Waveguide & \EO & 2.4 dB & \NA & \NA \\ \hline
                Lira et. al.~\cite{lira_electrically_2012} & 2012 & Waveguide & \EO & 3 dB & 70 dB & 0.04 \\ \hline
                Doerr et. al.~\cite{doerr_optical_2011}& 2011 & Waveguide & \EO & 2 dB & 11 dB & 0.18 \\ \hline
                Bhandare et. al.~\cite{bhandare_novel_2005}& 2005 & Waveguide & \EO & 30 dB & 8 dB & 3.75 \\ \hline
                
			    \hline \hline
                Cheng et. al.~\cite{cheng_terahertz-bandwidth_2025} & 2025 & Waveguide & \AO & 28 dB & 8 dB & 3.5 \\ 
				&  &  & \green & 12 dB & 2 dB & 6 \\ \hline
                 
                Sohn et. al.~\cite{sohn_electrically_2021} & 2021 & Resonator & \AO & 39.31 dB & 0.65 dB & \textbf{60.46} \\ 
				 &  &  & \green & 12.75 dB & 1.13 dB & 11.28 \\ \hline
                 
                Tian et. al.~\cite{tian_magnetic-free_2021} & 2021 & Resonator & \AO & 9.3 dB & 0.8 dB & 11.62 \\ 
				 &  &  & \green & 41 dB & 1.9 dB & 21.57 \\ \hline

			\end{tabular}
			\end{adjustwidth}

			\label{tab:FoM_table}

		\end{table}
	\end{centering}


\vspace{2cm}




{\footnotesize \begin{thebibliography}{10}
\expandafter\ifx\csname url\endcsname\relax
  \def\url#1{\texttt{#1}}\fi
\expandafter\ifx\csname urlprefix\endcsname\relax\def\urlprefix{URL }\fi
\providecommand{\bibinfo}[2]{#2}
\providecommand{\eprint}[2][]{\url{#2}}

\bibitem{bi_-chip_2011}
\bibinfo{author}{Bi, L.} \emph{et~al.}
\newblock \bibinfo{title}{On-chip optical isolation in monolithically integrated non-reciprocal optical resonators}.
\newblock \emph{\bibinfo{journal}{Nat. Photonics}} \textbf{\bibinfo{volume}{5}}, \bibinfo{pages}{758} (\bibinfo{year}{2011}).
\newblock \urlprefix\url{https://www.nature.com/articles/nphoton.2011.270}.

\bibitem{ghosh_ceyigsilicon--insulator_2012}
\bibinfo{author}{Ghosh, S.} \emph{et~al.}
\newblock \bibinfo{title}{Ce:{YIG}/{Silicon}-on-{Insulator} waveguide optical isolator realized by adhesive bonding}.
\newblock \emph{\bibinfo{journal}{Optics Express}} \textbf{\bibinfo{volume}{20}}, \bibinfo{pages}{1839} (\bibinfo{year}{2012}).
\newblock \urlprefix\url{https://opg.optica.org/oe/abstract.cfm?uri=oe-20-2-1839}.

\bibitem{huang_dynamically_2017}
\bibinfo{author}{Huang, D.} \emph{et~al.}
\newblock \bibinfo{title}{Dynamically reconfigurable integrated optical circulators}.
\newblock \emph{\bibinfo{journal}{Optica}} \textbf{\bibinfo{volume}{4}}, \bibinfo{pages}{23} (\bibinfo{year}{2017}).
\newblock \urlprefix\url{https://opg.optica.org/abstract.cfm?URI=optica-4-1-23}.

\bibitem{zhang_monolithic_2019}
\bibinfo{author}{Zhang, Y.} \emph{et~al.}
\newblock \bibinfo{title}{Monolithic integration of broadband optical isolators for polarization-diverse silicon photonics}.
\newblock \emph{\bibinfo{journal}{Optica}} \textbf{\bibinfo{volume}{6}}, \bibinfo{pages}{473} (\bibinfo{year}{2019}).
\newblock \urlprefix\url{https://opg.optica.org/abstract.cfm?URI=optica-6-4-473}.

\bibitem{du_monolithic_2018}
\bibinfo{author}{Du, Q.} \emph{et~al.}
\newblock \bibinfo{title}{Monolithic {On}-chip {Magneto}-optical {Isolator} with 3 {dB} {Insertion} {Loss} and 40 {dB} {Isolation} {Ratio}}.
\newblock \emph{\bibinfo{journal}{ACS Photonics}} \textbf{\bibinfo{volume}{5}}, \bibinfo{pages}{5010--5016} (\bibinfo{year}{2018}).
\newblock \urlprefix\url{https://pubs.acs.org/doi/10.1021/acsphotonics.8b01257}.

\bibitem{yan_waveguide-integrated_2020}
\bibinfo{author}{Yan, W.} \emph{et~al.}
\newblock \bibinfo{title}{Waveguide-integrated high-performance magneto-optical isolators and circulators on silicon nitride platforms}.
\newblock \emph{\bibinfo{journal}{Optica}} \textbf{\bibinfo{volume}{7}}, \bibinfo{pages}{1555} (\bibinfo{year}{2020}).
\newblock \urlprefix\url{https://opg.optica.org/abstract.cfm?URI=optica-7-11-1555}.

\bibitem{yan_ultra-broadband_2024}
\bibinfo{author}{Yan, W.} \emph{et~al.}
\newblock \bibinfo{title}{Ultra-broadband magneto-optical isolators and circulators on a silicon nitride photonics platform}.
\newblock \emph{\bibinfo{journal}{Optica}} \textbf{\bibinfo{volume}{11}}, \bibinfo{pages}{376} (\bibinfo{year}{2024}).
\newblock \urlprefix\url{https://opg.optica.org/abstract.cfm?URI=optica-11-3-376}.

\bibitem{hwang_all-fiber-optic_1997}
\bibinfo{author}{Hwang, I.~K.}, \bibinfo{author}{Yun, S.~H.} \& \bibinfo{author}{Kim, B.~Y.}
\newblock \bibinfo{title}{All-fiber-optic nonreciprocal modulator}.
\newblock \emph{\bibinfo{journal}{Optics Letters}} \textbf{\bibinfo{volume}{22}}, \bibinfo{pages}{507} (\bibinfo{year}{1997}).
\newblock \urlprefix\url{https://opg.optica.org/abstract.cfm?URI=ol-22-8-507}.

\bibitem{sohn_time-reversal_2018}
\bibinfo{author}{Sohn, D.~B.}, \bibinfo{author}{Kim, S.} \& \bibinfo{author}{Bahl, G.}
\newblock \bibinfo{title}{Time-reversal symmetry breaking with acoustic pumping of nanophotonic circuits}.
\newblock \emph{\bibinfo{journal}{Nature Photonics}} \textbf{\bibinfo{volume}{12}}, \bibinfo{pages}{91--97} (\bibinfo{year}{2018}).
\newblock \urlprefix\url{https://www.nature.com/articles/s41566-017-0075-2}.

\bibitem{kittlaus_electrically_2020}
\bibinfo{author}{Kittlaus, E.~A.} \emph{et~al.}
\newblock \bibinfo{title}{Electrically driven acousto-optics and broadband non-reciprocity in silicon photonics}.
\newblock \emph{\bibinfo{journal}{Nature Photonics}} \textbf{\bibinfo{volume}{15}}, \bibinfo{pages}{43--52} (\bibinfo{year}{2020}).
\newblock \urlprefix\url{https://www.nature.com/articles/s41566-020-00711-9}.

\bibitem{sohn_electrically_2021}
\bibinfo{author}{Sohn, D.~B.}, \bibinfo{author}{Örsel, O.~E.} \& \bibinfo{author}{Bahl, G.}
\newblock \bibinfo{title}{Electrically driven optical isolation through phonon-mediated photonic Autler--Townes splitting}.
\newblock \emph{\bibinfo{journal}{Nature Photonics}} \textbf{\bibinfo{volume}{15}}, \bibinfo{pages}{822--827} (\bibinfo{year}{2021}).
\newblock \urlprefix\url{https://www.nature.com/articles/s41566-021-00884-x}.

\bibitem{tian_magnetic-free_2021}
\bibinfo{author}{Tian, H.} \emph{et~al.}
\newblock \bibinfo{title}{Magnetic-free silicon nitride integrated optical isolator}.
\newblock \emph{\bibinfo{journal}{Nature Photonics}} \textbf{\bibinfo{volume}{15}}, \bibinfo{pages}{828--836} (\bibinfo{year}{2021}).
\newblock \urlprefix\url{https://www.nature.com/articles/s41566-021-00882-z}.

\bibitem{cheng_terahertz-bandwidth_2025}
\bibinfo{author}{Cheng, H.} \emph{et~al.}
\newblock \bibinfo{title}{A terahertz-bandwidth non-magnetic isolator}.
\newblock \emph{\bibinfo{journal}{Nature Photonics}}  (\bibinfo{year}{2025}).
\newblock \urlprefix\url{https://www.nature.com/articles/s41566-025-01663-8}.

\bibitem{bhandare_novel_2005}
\bibinfo{author}{Bhandare, S.} \emph{et~al.}
\newblock \bibinfo{title}{Novel nonmagnetic 30-{dB} traveling-wave single-sideband optical isolator integrated in {III}/{V} material}.
\newblock \emph{\bibinfo{journal}{IEEE Journal of Selected Topics in Quantum Electronics}} \textbf{\bibinfo{volume}{11}}, \bibinfo{pages}{417--421} (\bibinfo{year}{2005}).
\newblock \urlprefix\url{http://ieeexplore.ieee.org/document/1425478/}.

\bibitem{doerr_optical_2011}
\bibinfo{author}{Doerr, C.~R.}, \bibinfo{author}{Dupuis, N.} \& \bibinfo{author}{Zhang, L.}
\newblock \bibinfo{title}{Optical isolator using two tandem phase modulators}.
\newblock \emph{\bibinfo{journal}{Optics Letters}} \textbf{\bibinfo{volume}{36}}, \bibinfo{pages}{4293} (\bibinfo{year}{2011}).
\newblock \urlprefix\url{https://opg.optica.org/abstract.cfm?URI=ol-36-21-4293}.

\bibitem{lira_electrically_2012}
\bibinfo{author}{Lira, H.}, \bibinfo{author}{Yu, Z.}, \bibinfo{author}{Fan, S.} \& \bibinfo{author}{Lipson, M.}
\newblock \bibinfo{title}{Electrically {Driven} {Nonreciprocity} {Induced} by {Interband} {Photonic} {Transition} on a {Silicon} {Chip}}.
\newblock \emph{\bibinfo{journal}{Physical Review Letters}} \textbf{\bibinfo{volume}{109}}, \bibinfo{pages}{033901} (\bibinfo{year}{2012}).
\newblock \urlprefix\url{https://link.aps.org/doi/10.1103/PhysRevLett.109.033901}.

\bibitem{tzuang_non-reciprocal_2014}
\bibinfo{author}{Tzuang, L.~D.}, \bibinfo{author}{Fang, K.}, \bibinfo{author}{Nussenzveig, P.}, \bibinfo{author}{Fan, S.} \& \bibinfo{author}{Lipson, M.}
\newblock \bibinfo{title}{Non-reciprocal phase shift induced by an effective magnetic flux for light}.
\newblock \emph{\bibinfo{journal}{Nature Photonics}} \textbf{\bibinfo{volume}{8}}, \bibinfo{pages}{701--705} (\bibinfo{year}{2014}).
\newblock \urlprefix\url{https://www.nature.com/articles/nphoton.2014.177}.

\bibitem{doerr_silicon_2014}
\bibinfo{author}{Doerr, C.~R.}, \bibinfo{author}{Chen, L.} \& \bibinfo{author}{Vermeulen, D.}
\newblock \bibinfo{title}{Silicon photonics broadband modulation-based isolator}.
\newblock \emph{\bibinfo{journal}{Optics Express}} \textbf{\bibinfo{volume}{22}}, \bibinfo{pages}{4493} (\bibinfo{year}{2014}).
\newblock \urlprefix\url{https://opg.optica.org/oe/abstract.cfm?uri=oe-22-4-4493}.

\bibitem{dostart_optical_2021}
\bibinfo{author}{Dostart, N.}, \bibinfo{author}{Gevorgyan, H.}, \bibinfo{author}{Onural, D.} \& \bibinfo{author}{Popoviƒá, M.~A.}
\newblock \bibinfo{title}{Optical isolation using microring modulators}.
\newblock \emph{\bibinfo{journal}{Optics Letters}} \textbf{\bibinfo{volume}{46}}, \bibinfo{pages}{460} (\bibinfo{year}{2021}).
\newblock \urlprefix\url{https://opg.optica.org/abstract.cfm?URI=ol-46-3-460}.

\bibitem{yu_integrated_2023}
\bibinfo{author}{Yu, M.} \emph{et~al.}
\newblock \bibinfo{title}{Integrated electro-optic isolator on thin-film lithium niobate}.
\newblock \emph{\bibinfo{journal}{Nature Photonics}} \textbf{\bibinfo{volume}{17}}, \bibinfo{pages}{666--671} (\bibinfo{year}{2023}).
\newblock \urlprefix\url{https://www.nature.com/articles/s41566-023-01227-8}.

\bibitem{orsel_electro-optic_2023}
\bibinfo{author}{Örsel, O.~E.} \& \bibinfo{author}{Bahl, G.}
\newblock \bibinfo{title}{Electro-optic non-reciprocal polarization rotation in lithium niobate}.
\newblock \emph{\bibinfo{journal}{APL Photonics}} \textbf{\bibinfo{volume}{8}}, \bibinfo{pages}{096107} (\bibinfo{year}{2023}).
\newblock \urlprefix\url{https://pubs.aip.org/app/article/8/9/096107/2911672/Electro-optic-non-reciprocal-polarization-rotation}.

\bibitem{gao_thin-film_2024}
\bibinfo{author}{Gao, L.} \emph{et~al.}
\newblock \bibinfo{title}{Thin-film lithium niobate electro-optic isolator fabricated by photolithography assisted chemo-mechanical etching}.
\newblock \emph{\bibinfo{journal}{Optics Letters}} \textbf{\bibinfo{volume}{49}}, \bibinfo{pages}{614} (\bibinfo{year}{2024}).
\newblock \urlprefix\url{https://opg.optica.org/abstract.cfm?URI=ol-49-3-614}.

\bibitem{orsel_giant_2025}
\bibinfo{author}{Örsel, O.~E.} \emph{et~al.}
\newblock \bibinfo{title}{Giant {Nonreciprocity} and {Gyration} through {Modulation}-{Induced} {Hatano}-{Nelson} {Coupling} in {Integrated} {Photonics}}.
\newblock \emph{\bibinfo{journal}{Physical Review Letters}} \textbf{\bibinfo{volume}{134}}, \bibinfo{pages}{153801} (\bibinfo{year}{2025}).
\newblock \urlprefix\url{https://link.aps.org/doi/10.1103/PhysRevLett.134.153801}.

\bibitem{zhang_electronically_2019}
\bibinfo{author}{Zhang, M.} \emph{et~al.}
\newblock \bibinfo{title}{Electronically programmable photonic molecule}.
\newblock \emph{\bibinfo{journal}{Nature Photonics}} \textbf{\bibinfo{volume}{13}}, \bibinfo{pages}{36--40} (\bibinfo{year}{2019}).
\newblock \urlprefix\url{https://www.nature.com/articles/s41566-018-0317-y}.

\bibitem{peng_what_2014}
\bibinfo{author}{Peng, B.}, \bibinfo{author}{Özdemir, Ş.~K.}, \bibinfo{author}{Chen, W.}, \bibinfo{author}{Nori, F.} \& \bibinfo{author}{Yang, L.}
\newblock \bibinfo{title}{What is and what is not electromagnetically induced transparency in whispering-gallery microcavities}.
\newblock \emph{\bibinfo{journal}{Nature Communications}} \textbf{\bibinfo{volume}{5}}, \bibinfo{pages}{5082} (\bibinfo{year}{2014}).
\newblock \urlprefix\url{https://www.nature.com/articles/ncomms6082}.

\bibitem{kim_complete_2017}
\bibinfo{author}{Kim, J.}, \bibinfo{author}{Kim, S.} \& \bibinfo{author}{Bahl, G.}
\newblock \bibinfo{title}{Complete linear optical isolation at the microscale with ultralow loss}.
\newblock \emph{\bibinfo{journal}{Sci. Rep.}} \textbf{\bibinfo{volume}{7}}, \bibinfo{pages}{1647} (\bibinfo{year}{2017}).
\newblock \urlprefix\url{https://www.nature.com/articles/s41598-017-01494-w}.

\bibitem{shi_nonreciprocal_2018}
\bibinfo{author}{Shi, Y.}, \bibinfo{author}{Lin, Q.}, \bibinfo{author}{Minkov, M.} \& \bibinfo{author}{Fan, S.}
\newblock \bibinfo{title}{Nonreciprocal {Optical} {Dissipation} {Based} on {Direction}-{Dependent} {Rabi} {Splitting}}.
\newblock \emph{\bibinfo{journal}{IEEE Journal of Selected Topics in Quantum Electronics}} \textbf{\bibinfo{volume}{24}}, \bibinfo{pages}{1--7} (\bibinfo{year}{2018}).
\newblock \urlprefix\url{https://ieeexplore.ieee.org/document/8310902/}.

\bibitem{xu_mutual-coupling_2022}
\bibinfo{author}{Xu, Y.}, \bibinfo{author}{Liu, T.}, \bibinfo{author}{Liu, S.}, \bibinfo{author}{Sun, X.} \& \bibinfo{author}{Zhang, D.}
\newblock \bibinfo{title}{Mutual-{Coupling} in {High}-{Q} {Silicon} {Dual}-{Concentric} {Micro}-{Ring}/{Racetrack} {Resonator}}.
\newblock \emph{\bibinfo{journal}{IEEE Photonics Journal}} \textbf{\bibinfo{volume}{14}}, \bibinfo{pages}{1--7} (\bibinfo{year}{2022}).
\newblock \urlprefix\url{https://ieeexplore.ieee.org/document/9809804/}.

\bibitem{zhang_broadband_2022}
\bibinfo{author}{Zhang, Y.}, \bibinfo{author}{Zhong, K.}, \bibinfo{author}{Zhou, X.} \& \bibinfo{author}{Tsang, H.~K.}
\newblock \bibinfo{title}{Broadband high-{Q} multimode silicon concentric racetrack resonators for widely tunable {Raman} lasers}.
\newblock \emph{\bibinfo{journal}{Nature Communications}} \textbf{\bibinfo{volume}{13}}, \bibinfo{pages}{3534} (\bibinfo{year}{2022}).
\newblock \urlprefix\url{https://www.nature.com/articles/s41467-022-31244-0}.

\bibitem{weis_lithium_1985}
\bibinfo{author}{Weis, R.~S.} \& \bibinfo{author}{Gaylord, T.~K.}
\newblock \bibinfo{title}{Lithium niobate: {Summary} of physical properties and crystal structure}.
\newblock \emph{\bibinfo{journal}{Applied Physics A Solids and Surfaces}} \textbf{\bibinfo{volume}{37}}, \bibinfo{pages}{191--203} (\bibinfo{year}{1985}).
\newblock \urlprefix\url{http://link.springer.com/10.1007/BF00614817}.

\bibitem{desiatov_ultra-low-loss_2019}
\bibinfo{author}{Desiatov, B.}, \bibinfo{author}{Shams-Ansari, A.}, \bibinfo{author}{Zhang, M.}, \bibinfo{author}{Wang, C.} \& \bibinfo{author}{Lonƒçar, M.}
\newblock \bibinfo{title}{Ultra-low-loss integrated visible photonics using thin-film lithium niobate}.
\newblock \emph{\bibinfo{journal}{Optica}} \textbf{\bibinfo{volume}{6}}, \bibinfo{pages}{380} (\bibinfo{year}{2019}).
\newblock \urlprefix\url{https://opg.optica.org/abstract.cfm?URI=optica-6-3-380}.

\bibitem{tian_piezoelectric_2024}
\bibinfo{author}{Tian, H.} \emph{et~al.}
\newblock \bibinfo{title}{Piezoelectric actuation for integrated photonics}.
\newblock \emph{\bibinfo{journal}{Advances in Optics and Photonics}} \textbf{\bibinfo{volume}{16}}, \bibinfo{pages}{749} (\bibinfo{year}{2024}).
\newblock \urlprefix\url{https://opg.optica.org/abstract.cfm?URI=aop-16-4-749}.

\bibitem{dong_monolithic_2023}
\bibinfo{author}{Dong, Z.} \emph{et~al.}
\newblock \bibinfo{title}{Monolithic {Barium} {Titanate} {Modulators} on {Silicon}-on-{Insulator} {Substrates}}.
\newblock \emph{\bibinfo{journal}{ACS Photonics}} \textbf{\bibinfo{volume}{10}}, \bibinfo{pages}{4367--4376} (\bibinfo{year}{2023}).

\bibitem{demkov_ferroelectric_2024}
\bibinfo{author}{Demkov, A.~A.} \& \bibinfo{author}{Posadas, A.~B.}
\newblock \bibinfo{title}{Ferroelectric {BaTiO}$_{\textrm{3}}$ for {Electro}-{Optic} {Modulators} in {Si} {Photonics}}.
\newblock \emph{\bibinfo{journal}{IEEE Journal of Selected Topics in Quantum Electronics}} \textbf{\bibinfo{volume}{30}}, \bibinfo{pages}{1--13} (\bibinfo{year}{2024}).

\bibitem{mohl_bidirectional_2025}
\bibinfo{author}{Möhl, C.} \emph{et~al.}
\newblock \bibinfo{title}{Bidirectional {Microwave}-{Optical} {Conversion} with an {Integrated} {Soft}-{Ferroelectric} {Barium} {Titanate} {Transducer}}.
\newblock \emph{\bibinfo{journal}{Physical Review X}} \textbf{\bibinfo{volume}{15}}, \bibinfo{pages}{041044} (\bibinfo{year}{2025}).

\bibitem{kim_low_2025}
\bibinfo{author}{Kim, G.~I.}, \bibinfo{author}{Yim, J.} \& \bibinfo{author}{Bahl, G.}
\newblock \bibinfo{title}{Low loss monolithic barium titanate on insulator integrated photonics with intrinsic quality factor {\textgreater}1 million} (\bibinfo{year}{2025}).
\newblock \urlprefix\url{http://arxiv.org/abs/2507.17150}.
\newblock \bibinfo{note}{ArXiv:2507.17150 [physics]}.

\bibitem{poulton_coherent_2017}
\bibinfo{author}{Poulton, C.~V.} \emph{et~al.}
\newblock \bibinfo{title}{Coherent solid-state {LIDAR} with silicon photonic optical phased arrays}.
\newblock \emph{\bibinfo{journal}{Optics Letters}} \textbf{\bibinfo{volume}{42}}, \bibinfo{pages}{4091} (\bibinfo{year}{2017}).
\newblock \urlprefix\url{https://opg.optica.org/abstract.cfm?URI=ol-42-20-4091}.

\bibitem{spencer_optical-frequency_2018}
\bibinfo{author}{Spencer, D.~T.} \emph{et~al.}
\newblock \bibinfo{title}{An optical-frequency synthesizer using integrated photonics}.
\newblock \emph{\bibinfo{journal}{Nature}} \textbf{\bibinfo{volume}{557}}, \bibinfo{pages}{81--85} (\bibinfo{year}{2018}).
\newblock \urlprefix\url{https://www.nature.com/articles/s41586-018-0065-7}.

\bibitem{lucas_ultralow-noise_2020}
\bibinfo{author}{Lucas, E.} \emph{et~al.}
\newblock \bibinfo{title}{Ultralow-noise photonic microwave synthesis using a soliton microcomb-based transfer oscillator}.
\newblock \emph{\bibinfo{journal}{Nature Communications}} \textbf{\bibinfo{volume}{11}}, \bibinfo{pages}{374} (\bibinfo{year}{2020}).
\newblock \urlprefix\url{https://www.nature.com/articles/s41467-019-14059-4}.

\bibitem{knappe_microfabricated_2004}
\bibinfo{author}{Knappe, S.} \emph{et~al.}
\newblock \bibinfo{title}{A microfabricated atomic clock}.
\newblock \emph{\bibinfo{journal}{Applied Physics Letters}} \textbf{\bibinfo{volume}{85}}, \bibinfo{pages}{1460--1462} (\bibinfo{year}{2004}).
\newblock \urlprefix\url{https://pubs.aip.org/apl/article/85/9/1460/116970/A-microfabricated-atomic-clock}.

\bibitem{hummon_photonic_2018}
\bibinfo{author}{Hummon, M.~T.} \emph{et~al.}
\newblock \bibinfo{title}{Photonic chip for laser stabilization to an atomic vapor with 10$^{\textrm{-11}}$ instability}.
\newblock \emph{\bibinfo{journal}{Optica}} \textbf{\bibinfo{volume}{5}}, \bibinfo{pages}{443} (\bibinfo{year}{2018}).
\newblock \urlprefix\url{https://opg.optica.org/abstract.cfm?URI=optica-5-4-443}.

\bibitem{newman_architecture_2019}
\bibinfo{author}{Newman, Z.~L.} \emph{et~al.}
\newblock \bibinfo{title}{Architecture for the photonic integration of an optical atomic clock}.
\newblock \emph{\bibinfo{journal}{Optica}} \textbf{\bibinfo{volume}{6}}, \bibinfo{pages}{680} (\bibinfo{year}{2019}).
\newblock \urlprefix\url{https://opg.optica.org/abstract.cfm?URI=optica-6-5-680}.

\bibitem{delhaye_optical_2007}
\bibinfo{author}{Del’Haye, P.} \emph{et~al.}
\newblock \bibinfo{title}{Optical frequency comb generation from a monolithic microresonator}.
\newblock \emph{\bibinfo{journal}{Nature}} \textbf{\bibinfo{volume}{450}}, \bibinfo{pages}{1214--1217} (\bibinfo{year}{2007}).
\newblock \urlprefix\url{https://www.nature.com/articles/nature06401}.


\end{thebibliography}}
\end{document}